\documentclass[pdflatex,sn-nature]{sn-jnl}

\usepackage{graphicx}%
\usepackage{multirow}%
\usepackage{amsmath,amssymb,amsfonts}%
\usepackage{amsthm}%
\usepackage{mathrsfs}%
\usepackage[title]{appendix}%
\usepackage{xcolor}%
\usepackage{textcomp}%
\usepackage{manyfoot}%
\usepackage{booktabs}%
\usepackage{algorithm}%
\usepackage{algorithmicx}%
\usepackage{algpseudocode}%
\usepackage{listings}%
\usepackage{caption}
\usepackage{bm}

\theoremstyle{thmstyleone}%
\theoremstyle{thmstyletwo}%

\theoremstyle{thmstylethree}%

\begin{document}

\title[General Intelligent Imaging and Uncertainty Quantification by Deterministic Diffusion Model]{General Intelligent Imaging and Uncertainty Quantification by Deterministic Diffusion Model}


\author*[1]{\fnm{Weiru} \sur{Fan}}\email{weiru\_fan@zju.edu.cn}

\author[1]{\fnm{Xiaobin} \sur{Tang}}\email{12036056@zju.edu.cn}

\author*[2]{\fnm{Yiyi} \sur{Liao}}\email{yiyi.liao@zju.edu.cn}

\author[1,3,4]{\fnm{Da-Wei} \sur{Wang}}\email{dwwang@zju.edu.cn}

\affil[1]{\orgdiv{Zhejiang Province Key Laboratory of Quantum Technology and Device, School of Physics, and State Key Laboratory for Extreme Photonics and Instrumentation}, \orgname{Zhejiang University}, \orgaddress{\city{Hangzhou}, \postcode{310027}, \state{Zhejiang Province}, \country{China}}}

\affil[2]{\orgdiv{College of Information Science and Electronic Engineering}, \orgname{Zhejiang University}, \orgaddress{\city{Hangzhou}, \postcode{310027}, \state{Zhejiang Province}, \country{China}}}

\affil[3]{\orgdiv{College of Optical Science and Engineering}, \orgname{Zhejiang University}, \orgaddress{\city{Hangzhou}, \postcode{310027}, \state{Zhejiang Province}, \country{China}}}

\affil[4]{\orgdiv{Hefei National Laboratory}, \orgaddress{\city{Hefei}, \postcode{230088}, \state{Anhui province}, \country{China}}}


\abstract{Computational imaging is crucial in many disciplines from autonomous driving to life sciences. However, traditional model-driven and iterative methods consume large computational power and lack scalability for imaging. Deep learning (DL) is effective in processing local-to-local patterns, but it struggles with handling universal global-to-local (nonlocal) patterns under current frameworks. To bridge this gap, we propose a novel DL framework that employs a progressive denoising strategy, named the deterministic diffusion model (DDM), to facilitate general computational imaging at a low cost. We experimentally demonstrate the efficient and faithful image reconstruction capabilities of DDM from nonlocal patterns, such as speckles from multimode fiber and intensity patterns of second harmonic generation, surpassing the capability of previous state-of-the-art DL algorithms. By embedding Bayesian inference into DDM, we establish a theoretical framework and provide experimental proof of its uncertainty quantification. This advancement ensures the predictive reliability of DDM, avoiding misjudgment in high-stakes scenarios. This versatile and integrable DDM framework can readily extend and improve the efficacy of existing DL-based imaging applications.}

\maketitle


Computational imaging (CI) has emerged as a transformative paradigm that transcends the limitations of traditional ones through the joint design of optical imaging and computational algorithms \cite{bhandari_computational_2022}. Recently, the integration of deep learning (DL) has propelled CI into new territories, advancing various CI tasks such as compressed sensing \cite{qiao_deep_2020,yang_dagan_2018}, super-resolution imaging \cite{ouyang_deep_2018,qiao_rationalized_2023,fang_deep_2021}, non-line-of-sight imaging \cite{metzler_deep-inverse_2020,faccio_non-line--sight_2020}, imaging through scattering media \cite{li_deep_2018,saunders_computational_2019,rahmani_multimode_2018,feng_neuws_2023,hu_unsupervised_2023} and single photon imaging \cite{bian_high-resolution_2023}. Despite the success, a fundamental and common problem of CI tasks is to recover local structures from non-local data, e.g., from the speckle of multiple scattering \cite{zhang_high_2021,xu_high_2023} or intensity patterns of nonlinear processes \cite{fan_second_2021}. Existing DL-based CI methods are ineffective in dealing with the global-to-local transferring, due to the inherent locality bias of artificial neural networks \cite{rahaman_spectral_2019,xu_frequency_2020}. Moreover, current DL-based CI algorithms often lack interpretability and face difficulties in effectively evaluating the uncertainty, leading to unreliable imaging \cite{monga_algorithm_2021,li_interpretable_2022}. \par 

To augment the capability of DL in CI tasks, a typical practice involves integrating innovative modules into an artificial neural network tailored for certain tasks \cite{brady_deep_2020}. Enhancements such as attention mechanisms \cite{guo_attention_2022} and deeper layers \cite{he_deep_2016} are often employed to improve the content-awareness. While such strategies may improve the performance of non-local data processing, these methods result in expensive computation \cite{zhang_high_2021}. Another line of methods replaces the per-pixel scalar loss functions with adversarial training \cite{lucas_generative_2019} and perceptual metrics \cite{johnson_perceptual_2016} for better results. However, reconfiguring the loss function is generally incompatible with the Bayesian inference-based uncertainty estimation \cite{gal_bayesian_2022,sun_deep_2021}. Reliable uncertainty quantification, which is critical for informed decision-making and precise predictions, typically requires negative log-likelihood for the loss function \cite{xue_reliable_2019,feng_deep-learning-based_2021}.\par

Inspired by the non-equilibrium statistical physics \cite{sohl-dickstein_deep_2015}, the diffusion probabilistic model (DPM) \cite{ho_denoising_2020} has emerged as a powerful generative model for learning a target data distribution. Its flexibility and tractability have led to its widespread applications in multi-modal generation \cite{yang_diffusion_2023}, computer vision \cite{saharia_palette_2022}, life science \cite{luo_antigen-specific_2022} and material design \cite{xie_crystal_2022}. DPM breaks down the challenging generation task into many easier ones by modeling the image-to-image translation process as a Markov chain, yielding promise in solving the nonlocal CI tasks \cite{song_score-based_2021}. Although excellent performance is received on several key areas, DPM faces a prominent hurdle in CI that requires precise and instant processing \cite{song_denoising_2022,croitoru_diffusion_2023}. This challenge stems from the inherent uncertainty of the probability-based generative models, as well as the substantial computational resources required to produce superior outputs. On the other hand, the unpredictable execution path of DPM hinders the implementation of reliable Bayesian inference, impeding the assessment of prediction reliability in DL. This issue is particularly critical in high-stakes scenarios \cite{chen_concept_2020}, where DPM might generate plausible but false outputs, leading to misjudgments that are often only identified retrospectively.\par

\begin{figure}[htb]
\centering
\includegraphics[width=\textwidth]{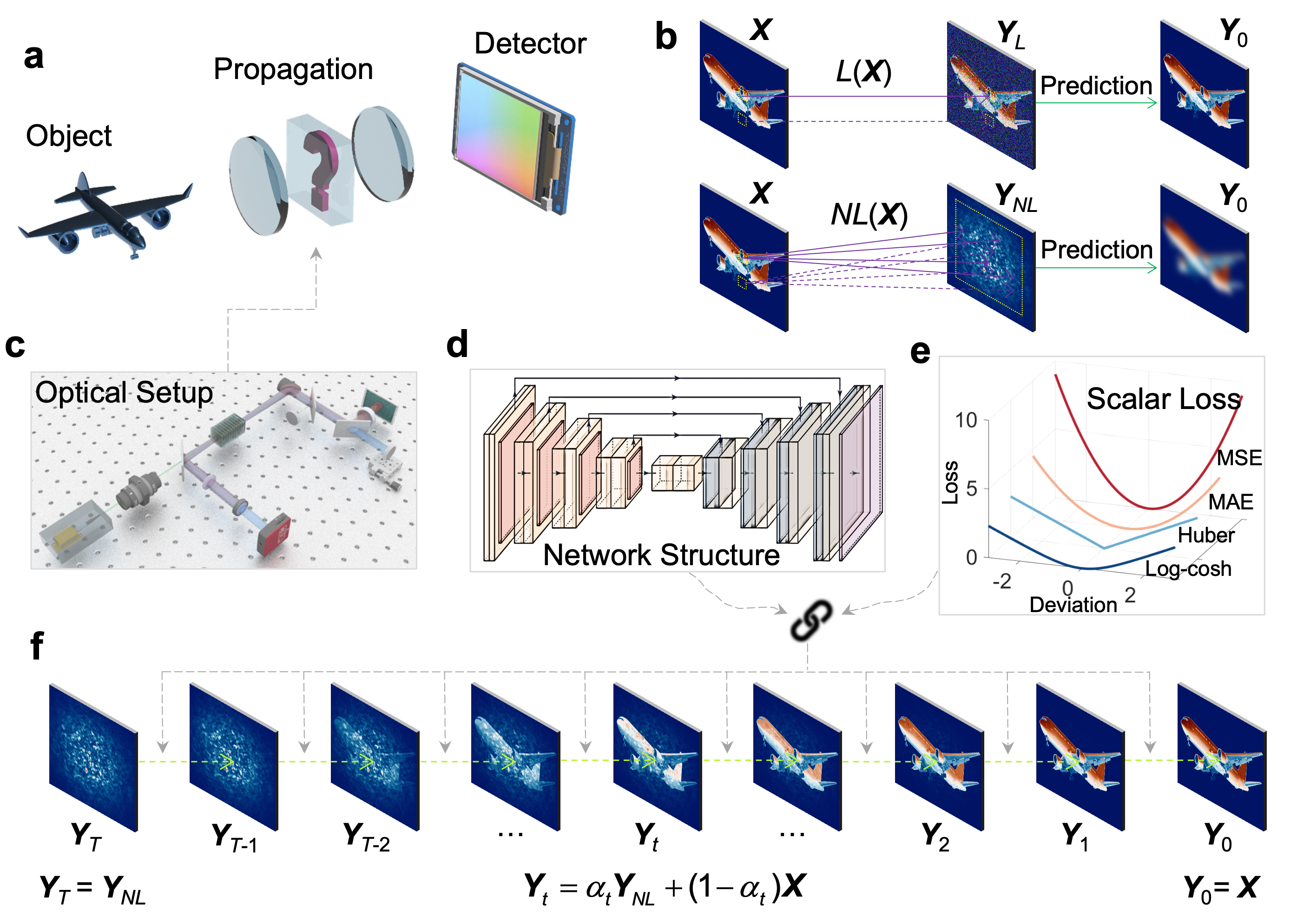}
\caption{DDM enabled CI. \textbf{a,} A schematics of optical imaging. The information of an object carried by light is transformed to the detector plane by optical components. The detector captures both the local and nonlocal patterns. \textbf{b,} The traditional DL-based imaging based on local and nonlocal patterns. Local patterns possess local similarity to the object, whereas nonlocal patterns have one-to-all correspondence. The DL image has a high fidelity for the local patterns but misses textures and details for the nonlocal patterns. \textbf{c, d, and e,} Three main sources of information loss in DL-based CI, which are optical setup, network structure, and scalar loss function. \textbf{f,} The structure of DDM. DDM utilizes a progressive denoise scheme to achieve the precise image reconstruction $\boldsymbol{Y}_0$ from $\boldsymbol{Y}_T$. The degraded image in each time step is obtained by the linear degeneration function of $\boldsymbol{Y}_T$ and $\boldsymbol{Y}_0$. $T$ is the number of the total degeneration steps. $\alpha_t$ is the predefined weight at time $t$. $L(\boldsymbol{X})$ and $\mathit{NL}(\boldsymbol{X})$ represent the local and nonlocal transformation for $\boldsymbol{X}$, respectively.}\label{fig1}
\end{figure} 

Here, we propose a deterministic diffusion model (DDM) which replaces the probabilistic Markov chain of DPM with a deterministic one, yielding a general and effective solution for CI tasks. The DDM eliminates the uncertainty of standard diffusion models while retaining their robust image translation capabilities. By using a linear combination of the detected image and the ground truth to construct the intermediate images in the diffusion process, we substantially reduced the required inference steps in DDM compared with previous diffusion models. This strategy significantly accelerates the inference process, showing the potential to enable real-time applications across various scenarios. Our experimental results demonstrate the DDM’s capability to reconstruct rich details and precise image profiles from nonlocal patterns, even for those generated by strong scattering and second harmonic generation. Moreover, the deterministic nature of DDM allows the employment of the Bayesian inference for quantifying the reliability and uncertainty of its output, facilitating the decision-making process in DL. The versatile DDM is applicable in most previous DL-based CI tasks, since it does not require building models that conform to real physical systems. The ‘plug and play’ characteristic enables DDM to act as an artificial general intelligence in CI, promising in remote sensing, medical diagnosis, reliability assessment, etc.\par


The traditional imaging mostly relies on lenses \cite{born_principles_2013}. General optical media such as random scattering media can disrupt the imaging relation and result in complex intensity patterns that possess no similarity to the original object (Fig. \ref{fig1}a). In general, these intensity patterns can be roughly divided into local ($\boldsymbol{Y}_L$) and nonlocal ($\boldsymbol{Y}_{\mathit{NL}}$) patterns. Local patterns maintain a morphological and structural similarity to the original object, whereas nonlocal patterns are featured by one-to-all correspondence between the object and the image (Fig. \ref{fig1}b). Local patterns characterized by centralized features in spatial correlation are more convenient for information processing. In most scenarios, DL utilizes local receptive fields (e.g., convolution) with shared weights to process large-scale image data from their local characteristics. However, the capability of DL significantly diminishes in perceiving and processing nonlocal features. The current DL methods often result in blurred profiles and distorted boundaries for nonlocal data (Fig. \ref{fig1}b).

The diminished efficacy of DL-based CI can be attributed to the information loss, where the smoothness prior \cite{bietti_inductive_2019,shalev-shwartz_understanding_2014} results in the loss of details in the image \cite{barbastathis_use_2019}. The information loss stems from three main sources: the optical setup (Fig. \ref{fig1}c), the network structure (Fig. \ref{fig1}d), and the scalar loss function (Fig. \ref{fig1}e). The optical loss is due to imperfect transformations of the optical elements and the mixing of the optical and electronic noise. While the neural network can extract the original signal from the complex background noise, it also loses information through irreversible data processing such as convolution and pooling \cite{liu_are_2020}. Furthermore, the output of the loss function is a single numerical value, which is insufficient to evaluate the pixel-level structural information \cite{johnson_perceptual_2016}, such that the neural network fails to capture the nuances of reconstructed images \cite{zhang_unreasonable_2018}. This challenge in evaluating content-awareness and texture in the training stage leads to detail-lacking images during inference. The information loss is particularly severe for nonlocal patterns, since structural features are lost in the optical transformation. In processing these nonlocal patterns in the neural network, the irreversible operations also introduce further information loss, consequently degrading the quality of image reconstruction.\par 

In this work, we propose DDM to substantially improve the imaging quality in CI (Fig. \ref{fig1}f). DDM employs a denoising method to progressively reconstruct high-fidelity images, similar to the principle of DPM but without the random degeneration \cite{ho_denoising_2020}. The DDM framework consists of the forward process and the reverse process. The forward process involves the stepwise degradation of the original object $\boldsymbol{X}$ (ground truth) into a pattern, including local and nonlocal patterns. The degraded image at a given time $t$ is obtained by a linear degradation function, which combines $\boldsymbol{Y}_T$ and $\boldsymbol{X}$($\boldsymbol{Y}_0$) with a weight $\alpha_t$. In the reverse process, neural networks and the linear degradation function are employed to infer $\boldsymbol{Y}_{t-1}$ from $\boldsymbol{Y}_t$ (see Methods for details). By iteratively performing these reverse diffusion steps, the final output of DDM is obtained by the culmination of this sequential prediction.\par 

In each step of the DDM process, we employ neural networks and scalar loss functions for training and inference, aiming to recover the original image from the degraded pattern. This process is equivalent to traditional single-step DL-based imaging when the number of degradation steps $T$ is set to 1. However, traditional DL approaches with single-step inference often fail to perceive nonlocal features, leading to significant information loss. In contrast, DDM employs multi-step inference to smoothly incorporate the nonlocal patterns into other desired patterns. At each step of DDM, $\boldsymbol{Y}_t$ is derived from $\boldsymbol{Y}_{t+1}$ by $\boldsymbol{Y}_t = \alpha_t \boldsymbol{Y}_T-\sum^T_{s=t+1} \left(\alpha_s-\alpha_{s-1}\right) R\left(\boldsymbol{Y}_s,s\right)$, where $R\left(\boldsymbol{Y}_s,s\right)$ is the neural network’s output for the input $\boldsymbol{Y}_s$ and $s$. Therefore, $\boldsymbol{Y}_t$ infuses the raw pattern $\boldsymbol{Y}_T$ and all the historical predicted images, effectively compensating for the lost information by reusing the raw pattern in each diffusion step. With the weight $\alpha_t$ decaying over time steps, the contribution of the raw pattern decreases until it reaches zero. As such, DDM enables high-quality imaging by effectively integrating information from nonlocal patterns throughout the diffusion process (Fig. \ref{fig1}f).\par 

\begin{figure}[tb]
\centering
\includegraphics[width=\textwidth]{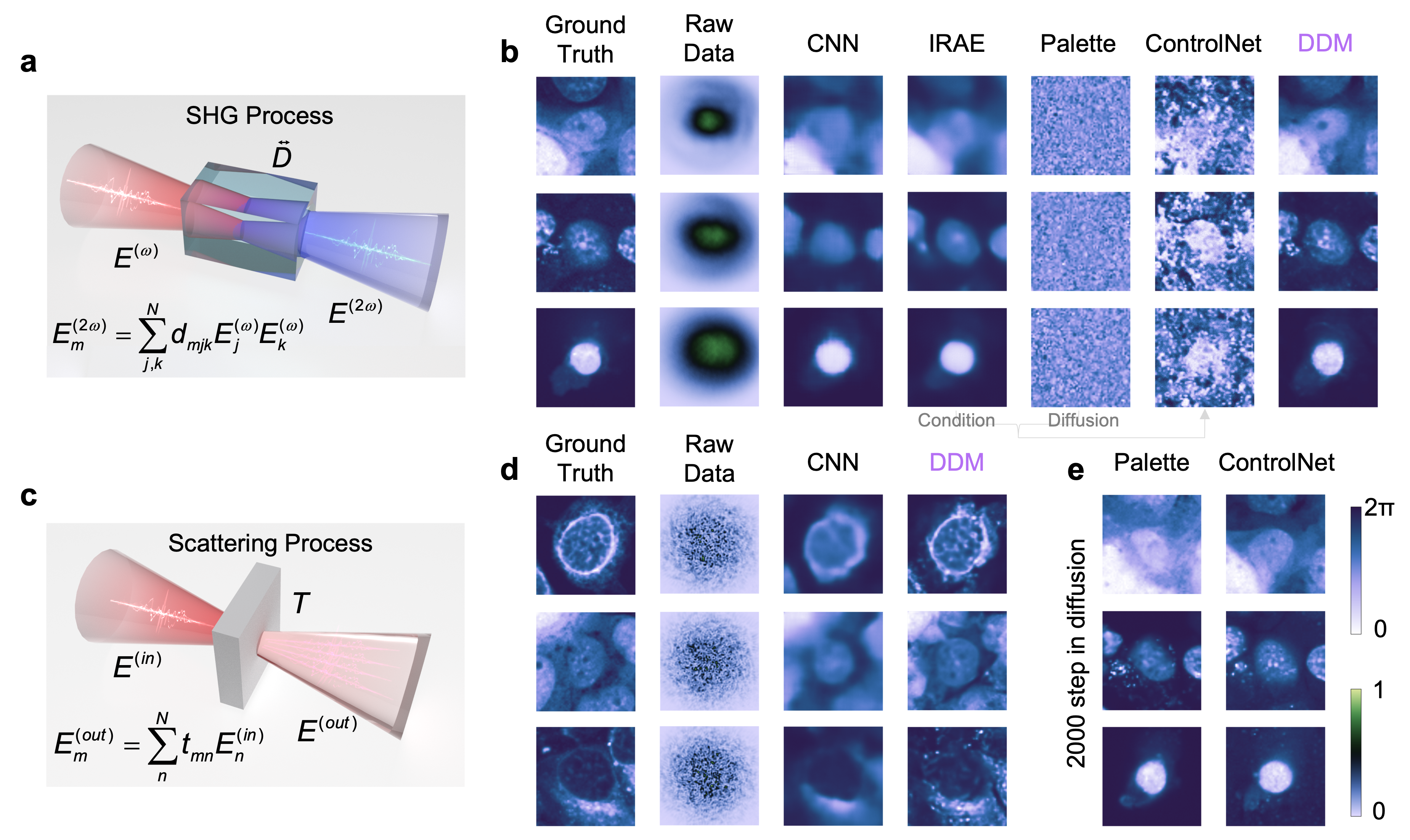}
\caption{The comparison of DDM with other approaches in imaging with nonlocal patterns. \textbf{a,} The schematic diagram of SHG, where the output field $\boldsymbol{E}^{(2\omega)}$ is generated from the input field $\boldsymbol{E}^{(\omega)}$ through a PPKTP nonlinear crystal. $\boldsymbol{E}^{(2\omega)}$ and $\boldsymbol{E}^{(\omega)}$ are related by a third-order tensor $\overleftrightarrow{\boldsymbol{D}}$ with the elements $d_{mjk}$ where $m$ labels the pixels of the output field and $j$ and $k$ label the pixels of the input field. \textbf{b,} The imaging results with different DL frameworks for the nonlocal raw pattern from SHG. CNN uses a Unet-like structure and the BiFormer module with a self-attention mechanism. ControlNet employs the Palette as the backbone of DPM and the result of IRAE as the condition input to control the output. \textbf{c,} The schematic diagram of strong scattering through a multimode fibre. The scattering process is represented by a transmission matrix $\boldsymbol{T}$ with the elements $t_{mn}$, which determines the relation between the output field $\boldsymbol{E}^{(out)}$ and the input field $\boldsymbol{E}^{(in)}$. \textbf{d,} The imaging results for the nonlocal raw pattern from scattering process. In \textbf{b} and \textbf{d}, the diffusion step number is 100 for all diffusion model. \textbf{e,} The DPM imaging in the SHG process with 2000 diffusion steps.}\label{fig2}
\end{figure}

To test the performance of DDM, we apply it to imaging through two representative optical processes: second harmonic generation (SHG) and strong scattering, both of which involve nonlocal patterns. For the SHG, the original image is encoded into the wavefront phase of a fundamental light field $\boldsymbol{E}^{(\omega)}$, and the SHG process transforms phase images to intensity images (Fig. \ref{fig2}a) \cite{moon_measuring_2023}. The detector records the SHG intensity pattern $\left|\boldsymbol{E}^{(2\omega)}\right|^2$ which exhibits a nonlocal spatial distribution, since each pixel in the output image is generated from all input pixels. To verify the advantage of DDM, we conduct comparison experiments with different DL frameworks on the identical dataset. These results are shown in Fig. \ref{fig2}b, demonstrating a clear superiority of DDM in reconstructing images from nonlocal patterns.\par 

Here, our proposed DDM is compared against several representative DL-based CI approaches. We first apply the widely used convolutional neural networks (CNN) with the state-of-the-art nonlocal module \cite{zhu_biformer_2023} to reconstruct images from the SHG patterns. However, the numerous irreversible operations in CNNs lead to significant information loss in this single-step formation, resulting in distorted profiles in the reconstructed images. We then use the invertible restoring autoencoder (IRAE) \cite{liu_are_2020}, which possesses a lossless reversible structure. However, the IRAE can only produce the profile (low frequency components) of the ground truth due to the optical loss and the ineffective scalar loss function for nonlocal features. Furthermore, we evaluate two DPM-based approaches. One is an image-to-image conditional DPM called Palette \cite{saharia_palette_2022}, where we perform the conditional denoising by sampling the standard Gaussian noise, conditioned on the raw pattern. The ControlNet is further employed to refine the sample results \cite{zhang_adding_2023}. For a fair comparison, we conduct diffusion steps for Palette and ControlNet identical to those in DDM (100 steps in Fig. \ref{fig2}b). Note that both DPMs fail to obtain coherent images with relatively few diffusion steps. In contrast, the DDM successfully reconstructs images with rich details and precise profiles from the nonlocal raw pattern. Despite that DPM has the capability to yield seemingly true results when the diffusion step is large enough (Fig. \ref{fig2}e), these randomly generated images are inconsistent for various reverse diffusion processes. The inherent randomness renders DPM unreliable in imaging from nonlocal patterns with high uncertainty.

In DDM, we employ a deterministic degeneration function to degrade the ground truth into the corresponding nonlocal pattern, in contrast to DPM which degrades it into a Gaussian noise. Thanks to the deterministic design, DDM allows for reconstructing high-fidelity images from nonlocal patterns within a few diffusion steps. We also demonstrate the power of DDM in imaging through strong scattering media. The original image is still encoded in the wavefront phase of the input light field $\boldsymbol{E}^{(in)}$. After scattering each pixel of the output $\boldsymbol{E}^{(out)}$ contains scattered light from all pixels of the input (Fig. \ref{fig2}c) \cite{vellekoop_focusing_2007}. The detected nonlocal pattern $\left|\boldsymbol{E}^{(out)}\right|^2$ is fed into DDM. The results shown in Fig. \ref{fig2}d demonstrate the versatility of DDM in dealing with nonlocal patterns. \par 

The success of DDM, from a mathematical perspective, lies in its formation that progressively transforms the nonlocal patterns into local ones through the degeneration function. This function linearly combines the ground truth and raw pattern at the pixel level, effectively simplifying the task of the neural network at each step. Substantially, this procedure of DDM can be treated as a residual connection at the network level \cite{srivastava_highway_2015}. From a physics standpoint, the working principle of DDM is similar to the adiabatic process \cite{farhi_quantum_2001,roland_quantum_2002}. Each employment of the neural network recovers a small component of the image from the degraded patterns $\boldsymbol{Y}_t$. This divides the nonlocal pattern imaging into many sequential local pattern estimations, avoiding the information loss caused by predicting from non-local patterns in a single step. \par 

By removing randomness, the deterministic nature of DDM allows for precise quantification of the uncertainty in its predictions. This uncertainty quantification provides pixel-level confidence in the DDM’s prediction and offers insights into the quality of the model and the dataset \cite{sun_deep_2021}, potentially aiding in problem understanding and informed decision-making. To achieve this, Bayesian inference is introduced to capture the uncertainty. Here, a Bayesian neural network (BNN), which replaces the deterministic network weights with random variables, is used with the trainable dropout to perform Bayesian inference in deep Gaussian processes \cite{xue_reliable_2019,gal_concrete_2017}. During training, the negative log-likelihood serves as the loss function, enabling automatic capture of the uncertainty without requiring the ground truth uncertainty for learning (details in Method). During the inference process, the BNN simultaneously outputs the predicted results and the estimated uncertainties. \par 

\begin{figure}[hbt]
\centering
\includegraphics[width=\textwidth]{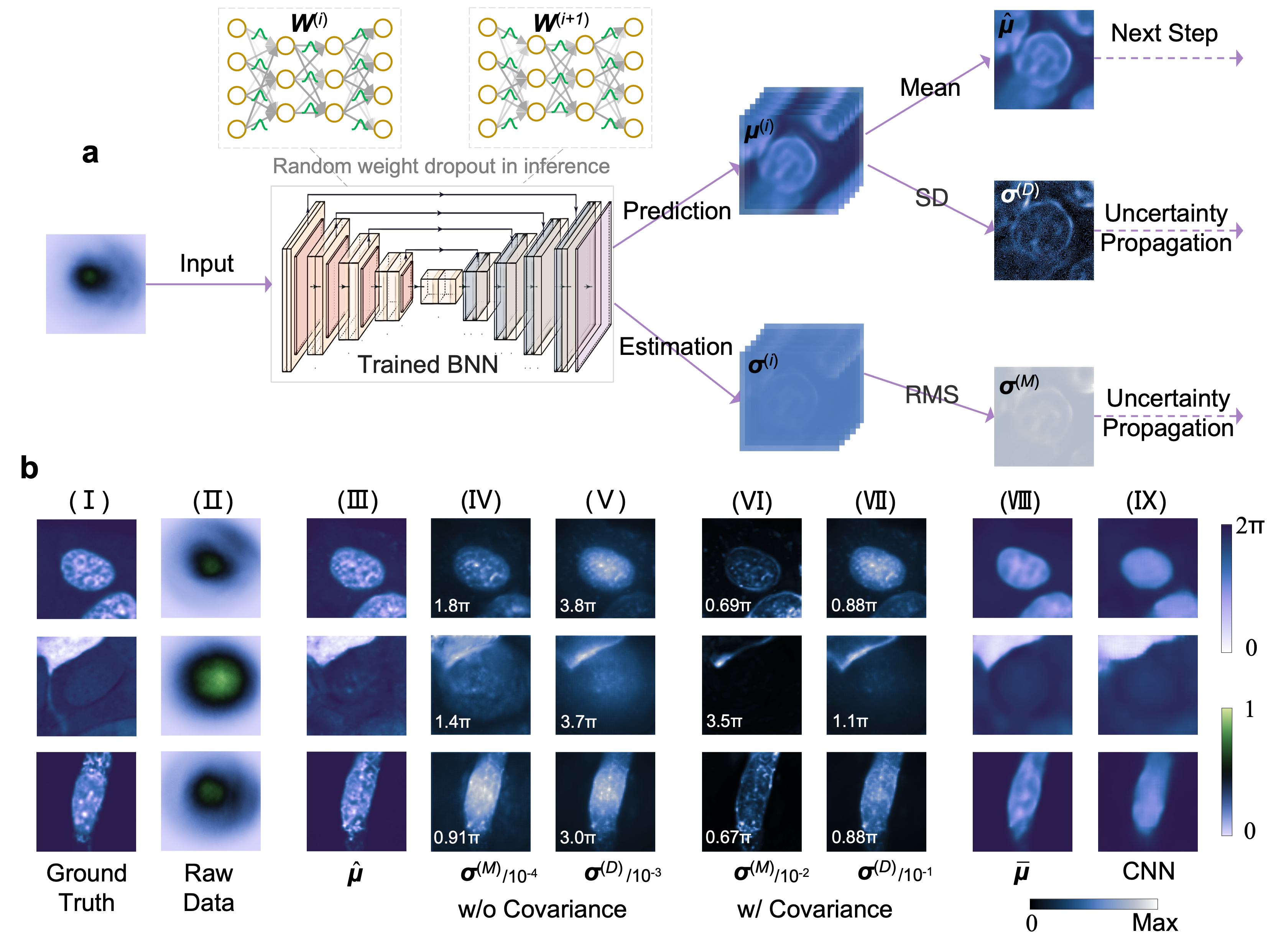}
\caption{Uncertainty quantification. \textbf{a,} Overview of one diffusion step for the uncertainty estimation. A trained BNN receives the input, and maintains activated dropout to predict the output, $\boldsymbol{\mu}^{(i)}$, and the standard deviation, $\boldsymbol{\sigma}^{(i)}$. The process is repeatedly performed to produce the set of $\boldsymbol{\mu}^{(i)}$ and $\boldsymbol{\sigma}^{(i)}$. The predictive mean is computed by averaging all samples of $\boldsymbol{\mu}^{(i)}$, and is regarded as the predicted result of the BNN in the current step. The model uncertainty is determined by the standard deviation (SD) of the predicted results $\boldsymbol{\mu}^{(i)}$, while the data uncertainty is quantified by calculating the root mean squared (RMS) of the estimated uncertainty $\boldsymbol{\sigma}^{(i)}$. \textbf{b,} The results of uncertainty quantification. The \uppercase\expandafter{\romannumeral 3}-\uppercase\expandafter{\romannumeral 5} column shows the results without the covariance in propagation of uncertainty, while the \uppercase\expandafter{\romannumeral 6} and \uppercase\expandafter{\romannumeral 7} columns are with the covariance. The \uppercase\expandafter{\romannumeral 8} column represents the most trusted results by averaging all samples and paths, yielding the contour of ground truth which is the same as the results of CNN.}\label{fig3}
\end{figure}

Activating dropout in the inference process, the BNN with random weights yields a series of predictions $\left\{\boldsymbol{\mu}^{(i)}\right\}$ and standard deviation (uncertainty) $\left\{\boldsymbol{\sigma}^{(i)}\right\}$ by repeatedly performing this procedure by $i$ times. Using Monte Carlo dropout, we can then calculate the various uncertainties (Fig. \ref{fig3}a). Generally, the two main types of uncertainty are particularly relevant for the users, data (epistemic) uncertainty $\boldsymbol{\sigma}^{(D)}$ and model (aleatoric) $\boldsymbol{\sigma}^{(M)}$ uncertainty \cite{gal_bayesian_2022}. Data uncertainty $\boldsymbol{\sigma}^{(D)}$ refers to the intrinsic uncertainty of a system and the observed data, such as the presence of noise, while model uncertainty $\boldsymbol{\sigma}^{(M)}$ reflects imperfections of the models caused by factors like model error or incomplete training. In each diffusion step, both model and data uncertainty can be quantified by analyzing the predicted output $\boldsymbol{\mu}^{(i)}$ and the estimated standard deviation $\boldsymbol{\sigma}^{(i)}$, respectively. The predictive mean is derived by averaging all predicted outputs (Fig. \ref{fig3}a). This mean is considered as a result of the current step and harnessed to generate the input for the next diffusion step by the deterministic degeneration function.\par

Because of the multistep operation, determining the total uncertainty requires the propagation of uncertainty. When naively assuming the input and output are mutually independent, the propagation of uncertainty can be simply calculated by the formula $\sum_{t=1}^T \left(\alpha_t-\alpha_{t-1}\right) \boldsymbol{\sigma}^{(M/D)}_t$. However, as depicted in Fig. \ref{fig3}\uppercase\expandafter{\romannumeral 4} and \ref{fig3}\uppercase\expandafter{\romannumeral 5}, it results in an unrealistic uncertainty that is less than the minimum discrete interval of the experimental instruments. This is because the naive assumption does not hold in practice, since the input and output are dependent on each other. Therefore, the propagation of uncertainty must incorporate a term that considers the covariance between these variables. However, calculating covariance is intractable due to the absence of analytical equations, as the input and output are associated through a `black box' neural network.\par 

To address this challenge, we introduce the Monte Carlo simulation to estimate the covariance, enabling effective uncertainty quantification (see Method for detailed derivation). When considering the covariance in the propagation of uncertainty, the uncertainty maps can be accurately obtained as shown in Fig. \ref{fig3}\uppercase\expandafter{\romannumeral 6} and \ref{fig3}\uppercase\expandafter{\romannumeral 7}. These results indicate that higher uncertainty usually appears at the edge and sharp regions of objects, while the model uncertainty is smaller than the data uncertainty by approximately one order of magnitudes. The reason is that DDM utilizes the degeneration function to perform sampling training, which is essentially equivalent to data augmentation. With $T$ diffusion steps, the amount of training data is equal to $T$-fold increase of the original dataset. This not only reduces the model uncertainty, but also mitigates the risk of overfitting. Additionally, thanks to random weights of BNN, DDM can regain multiple paths to reconstruct images from nonlocal patterns. Particularly, the most trusted path can be obtained by averaging throughout all sampled paths (Fig. \ref{fig3}\uppercase\expandafter{\romannumeral 8}). The most trusted results reduce high-frequency components, while accurately maintaining the contour. This is similar to the results of CNN (Fig. \ref{fig3}\uppercase\expandafter{\romannumeral 9}), but with more precise structure and boundary.

There is a counterintuitive observation where DDM’s pixel-wise quantitative metrics can be lower than those of single-step frameworks, yet its results are more visually authentic and of higher quality in terms of perceptual metrics. The discrepancy arises as the pixel-wise index is inadequate for capturing perceptual differences between reconstructed images and ground-truth \cite{johnson_perceptual_2016, zhang_unreasonable_2018}. For example, an overly smoothed image may have a low pixel-wise mean squared error while being undesirable in CI. To faithfully measure the image quality, the perceptual metrics must be considered for quantitative evaluation \cite{Sheikh_image_2006}. We compare these two types of metrics and demonstrate opposite behavior for pixel-wise and visual perception indexes, shown in Extended Data Fig. \ref{figed1}. Future explorations may further enhance the pixel-wise metrics through more advanced training procedures and cutting-edge architectures. For instance, a promising avenue is to embed scheduling ($\alpha_t$) into the neural network as learnable parameters to search for the optimal weighted schedule \cite{nichol_improved_2021}. This enables an adaptive balance of the results at each diffusion step, $\boldsymbol{Y}_0 = \alpha_t \boldsymbol{Y}_T-\sum_{s=1}^T\left(\alpha_s-\alpha_{s-1}\right) R\left(\boldsymbol{Y}_s,s\right)$.\par 

In addition, since the diffusion of DDM is a deterministic process and $\boldsymbol{Y}_0$ can be inferred at any step, it offers the flexibility to use any number of diffusion steps for inference. Specifically, the diffusion step $T$ controls the performance of DDM, where a larger $T$ tends to provide more details and a smaller $T$ leads to more confident results. The effectiveness of image reconstruction with DDM is the intrinsic trade-off between the richness of details and the confidence of results. Moreover, the deterministic diffusion stabilizes training compared to the random diffusion and can accelerate convergence. These advantages facilitate the development of high-performance models with reduced time cost and computing resources.\par 

As a versatile tool, DDM is not only capable of global-to-local imaging but can also be applied to local-to-local data. The flexibility of DDM extends further with appropriate modifications to its degeneration function, allowing it to process diverse data types, including voice, video, and 3D point clouds. When the degeneration function matches an accurate physical model, DDM can be used to study dynamic processes in the corresponding systems. Meanwhile, DDM can incorporate latent diffusion \cite{rombach_high-resolution_2022} for perceptual compression, accelerating training and reducing computational complexity. The model can also benefit from replacing traditional CNN backbones with transformers equipped with self-attention mechanisms \cite{peebles_scalable_2023}, further boosting its effectiveness and expanding its applicability.\par 

In conclusion, we propose the deterministic diffusion model (DDM), marking a significant advancement in DL-based CI. DDM employs multi-step processing to tackle challenging image reconstruction tasks, particularly effective for global-to-local transferring. It stands out from previous state-of-the-art frameworks by reconstructing images with rich details and precise profiles from nonlocal patterns, while other end-to-end methods typically yield blurred profiles and distorted boundaries. We experimentally demonstrate the superior imaging performance of DDM with nonlocal patterns generated by SHG and strong scattering. Moreover, we present a theoretical framework and experimental proof that uncertainty can be captured by introducing Bayesian inference into DDM. This multi-step image reconstruction significantly reduces model uncertainty, thereby robustly solving CI problems and offering generalizability across different samples. With its `plug and play' capability and high compatibility, the DDM framework is a versatile tool potentially applicable to a wide array of promising DL-based applications, including data engineering, imaging under extreme scenarios, and critical assessments.

\section*{Methods}\label{secm}
\subsection*{Experimental setup for generating nonlocal pattern}\label{secm1}
For the nonlinear SHG configuration, an infrared pulsed laser is used as a fundamental field (1064 nm, 330 fs; pump: SPIRIT 1040-16\_30-HE; OPA: Orpheus-HP, Spectra-Physics). The laser beam passes through a bandpass filter (FLH1064-10, Thorlabs) with full width at half maximum of 10 nm at a central wavelength of 1064 nm. The laser spot is expanded with a 6 mm diameter by a telescope system formed by a pair of lenses and is shaped by placing an iris in the confocal plane of the lens pair. An SLM (X13139-09, Hamamatsu) is used to modulate the wavefront phase of the laser beam. The modulated light is then collected and focused into a nonlinear crystal (PPKTP with 2 cm length; periodically poled potassium titanyl phosphate, Raicol) by a lens with focal length of 275mm to generate the SHG signal. The SHG signal is filtered by a 532 nm bandpass filter (FLH05532-10, Thorlabs), and the intensity pattern is recorded by a CCD (Prosilica GT1910, AVT) with 512 $\times$ 512 pixels and 8-bit grayscale without an imaging lens. The modulation region has 256 $\times$ 256 pixels on the SLM (the area for each pixel is 12.5 $\times$ 12.5 $\mathrm{\mu m}^2$), which corresponds to the central position of laser spot. The crystal is working at 30 $^{\circ}\rm{C}$ temperature. The detailed experimental setup is shown in Extended Data Fig. \ref{figed2}a. \par 

In the linear strong scattering experiment, the semiconductor laser (Precilaser, YFA-SF-1064-50-CW) is used to provide the coherent continuous wave with 532 nm wavelength, and we perform the expanding and shaping processes for the light resource. Then the collimated laser beam illuminates onto the SLM in which a phase map is imposed into the incident light. The modulated light is subsequently and focused into a multimode fiber (MMF; M15L02, Thorlabs) by an objective (10$\times$/0.3NA, Olympus). Owing to defects, bending, mode coupling, mode superposition and dispersion between modes, a nonlocal speckle pattern is generated when a coherent beam goes through the MMF. Another objective (20$\times$/0.4NA, Olympus) is used at the output end of MMF to collect its speckles and record them by a CCD. The modulation size, recorded pattern resolution and spot size of the unmodulated light are the same as in the SHG experiment. The MMF is bent into a circle ring with a diameter of about 8 cm to enhance the scattering (see Extended Data Fig. \ref{figed2}b for the experimental setup).\par 

\subsection*{Dataset preparation}\label{secm2}
The phase maps are prepared by the Cells Out Of Sample 7-class (COOS-7) dataset \cite{lu_cells_2019}, which contains 132,209 crops of mouse cells. In this dataset, each cell image is stained with one of seven fluorescent proteins which are the endoplasmic reticulum, inner mitochondrial membrane, Golgi, peroxisomes, early endosome, cytosol, and nuclear envelope. The distribution of these fluorescent proteins is significantly different in cells, enabling the complex structure. The original dataset has two channels for each image, representing the protein and nucleus. We directly stack the two channels by the additive operation and normalize the result to the interval [0,~1], which makes each pattern contain a more sophisticated structure. We merely employ the training and test4 dataset in the entire COOS-7 repository, and randomize their order to build a new image dataset with 72228 cell images. The bicubic interpolation is applied to upsample the original 64 $\times$ 64 pixels image to 256 $\times$ 256. The cell images are then imposed in sequence onto SLM to generate the corresponding raw pattern with 512 $\times$ 512, and perform downsampling for the raw pattern to obtain the 256 $\times$ 256 pattern. Some samples are shown in Extended Data Fig. \ref{figed3} for the experimental setup.\par 

\subsection*{CNN embedding the BiFormer module}\label{secm3}
The CNN is designed with a Unet-like structure \cite{ronneberger_u-net_2015} in our experiment, but the original convolution block is replaced by the BiFormer module to improve the ability of nonlocal processing. Specifically, upon the input of 256 $\times$ 256 raw patterns with a single channel, the 3 $\times$ 3 convolution is employed to extend the pattern to 64 channels with the same dimensions, and then transformed through network layers, including multiple BiFormer modules and convolution-based pooling in each layer (BiFormer with 2, 4, 2 and 1 time, respectively), to form deeper feature maps, defined as encoding. The feature maps then go through a decoding process (64 $\times$ 64 $\times$ 256, 128 $\times$ 128 $\times$ 128, 256 $\times$ 256 $\times$ 64, and 256 $\times$ 256 $\times$ 64 blocks, respectively), where layers are concatenated with the feature maps that have the same dimensions in the encoding process, operated by multiple BiFormer modules (2, 4 and 2 time, respectively) and transpose convolution excepting the 3 $\times$ 3 convolution in the last layer, which is symmetric with the encoding process. The final output is generated by the 1 $\times$ 1 convolution to transform the last feature maps with 256 $\times$ 256 $\times$ 64 to a single channel. The detailed structure for the BiFormer module can be found in Ref. \cite{zhu_biformer_2023}.

\subsection*{IRAE Implementation}\label{secm4}
For the view of mutual information, when a deep neural architecture is invertible, the information concerning the input can be completely retained \cite{liu_are_2020}. To achieve the invertibility, IRAE uses the invertible flow-based generative algorithms to build the deep network architecture. IRAE has an encoder-decoder structure resembling the Unet, but each layer is replaced by the invertible flow-based block \cite{liu_are_2020}. The encoder is constituted by $L_f$ flow-based block, where each block contains a squeeze, $K_f$ steps of the flow module and a split. Each step of the flow module has an activation normalization, an invertible 1 $\times$ 1 convolution and an affine coupling. The decoder has a symmetric structure with the encoder but each flow-based block is inverted, where data flow is similar to reversing through the encoder but without using the same weight. Experimentally, we set the hyperparameters as follows, $L_f$ = 4, $K_f$ = 16, and $W_f$ = 128, where $W_f$ is a parameter for the affine coupling.

\subsection*{The principal of conditional DPM (Palette)}\label{secm5}
Diffusion models involve a forward diffusion process and a reverse denoising process, corresponding to the training and generation. The forward diffusion process progressively imposes random Gaussian noise to a pattern $\boldsymbol{y}_0 = \boldsymbol{x}$, given samples from a data distribution $\boldsymbol{x}\sim q(\boldsymbol{x})$ over $T$ iterations (diffusion steps) according to a predefined variance schedule $\left\{\beta_1, \beta_2, \dots , \beta_T\right\}$. Owing to the notable property of Gaussian distribution, the forward process can be marginalized to sample $\boldsymbol{y}_t$ at an arbitrary timestep $t$ in a closed form \cite{ho_denoising_2020},
\begin{equation}
q\left(\boldsymbol{y}_t \mid \boldsymbol{y}_0\right)=\mathcal{N}\left(\boldsymbol{y}_t ; \sqrt{\gamma_t} \boldsymbol{y}_0,\left(1-\gamma_t\right) \mathbf{I}\right)
\label{equm1}
\end{equation}
where ${{\gamma }_{t}}=\sum\nolimits_{s=1}^{t}{(1-{{\beta }_{s}})}$. The posterior distribution of $\boldsymbol{y}_{t-1}$ given $(\boldsymbol{y}_0, \boldsymbol{y}_t)$ can be achieved by the Gaussian parameterization of the forward process with the same closed form formulation \cite{song_denoising_2022},
\begin{equation}
    q\left({{\boldsymbol{y}}_{t-1}}\mid{{\boldsymbol{y}}_{0}},{{\boldsymbol{y}}_{t}}\right)=\mathcal{N}\left({{\boldsymbol{y}}_{t-1}};\sqrt{{{\gamma }_{t-1}}}{{\boldsymbol{y}}_{0}}+\sqrt{1-{{\gamma }_{t-1}}-\sigma _{t}^{2}}\frac{{{\boldsymbol{y}}_{t}}-\sqrt{{{\gamma }_{t}}}{{\boldsymbol{y}}_{0}}}{\sqrt{1-{{\gamma }_{t}}}},\sigma _{t}^{2}\mathbf{I}\right)
    \label{equm2}
\end{equation}
where $\sigma_{t}^{2}=\eta \left[{\left( 1-{{\gamma }_{t-1}} \right)}/{\left( 1-{{\gamma }_{t}} \right)} \right]\left( 1-{{{\gamma }_{t}}}/{{{\gamma }_{t-1}}} \right)$ with $\eta \in [0,1]$. The form of the mean function is chosen in order to ensure that Eq. (\ref{equm1}) is satisfied at any time $t$. \par 

According to Eq. (\ref{equm1}) and by applying the reparameterization trick, we can generate the degraded noisy pattern $\boldsymbol{y}^*$ at any time $t$ as,
\begin{equation}
\boldsymbol{y}_t^*=\sqrt{\gamma_t} \boldsymbol{y}_0+\sqrt{1-\gamma_t} \boldsymbol{\varepsilon}, \boldsymbol{\varepsilon} \sim \mathcal{N}(0, \mathbf{I})
\label{equm3}
\end{equation}
The aim of DPM is to learn a reverse process $p_\theta(\boldsymbol{y}_0)$ that approximates $q(\boldsymbol{x})$ to invert the forward process by a parameterized neural network model $f_\theta\left(\boldsymbol{y}_c, \boldsymbol{y}_t^*, \gamma_t\right)$, where $\boldsymbol{y}_c$ is the condition (here it is the raw pattern) and $\theta$ is the weight of the network. The training involves feeding a condition $\boldsymbol{y}_c$, a degraded noisy pattern $\boldsymbol{y}_t^*$ and the current noise level $\gamma_t$ to learn the prediction of the noise vector $\boldsymbol{\varepsilon}$. The objective is maximizing a weighted variational lower-bound on the likelihood by optimizing the formulation \cite{saharia_palette_2022},
\begin{equation}
\mathbb{E}_{\boldsymbol{y}_c, \boldsymbol{y}_t^*} \mathbb{E}_{\boldsymbol{\varepsilon}, \gamma_t}\left\|f_\theta\left(\boldsymbol{y}_c, \boldsymbol{y}_t^*, \gamma_t\right)-\boldsymbol{\varepsilon}\right\|
\label{equm4}
\end{equation}\par 

Owing to the construction of the forward process, the prior distribution of the final state $\boldsymbol{y}_T$ is approximated as a standard normal distribution. This means that the sampling process can start from the pure Gaussian noise pattern and progressively denoise from the $\boldsymbol{y}_T$ over $T$ steps. Concretely, the neural network $f_\theta$ is trained to estimate the noise vector $\boldsymbol{\varepsilon}$ from a degraded noisy pattern $\boldsymbol{y}^*$ under the condition $\boldsymbol{y}_c$ and noise level $\gamma_t$. Given $\boldsymbol{y}_t^*$, we can estimate $\boldsymbol{y}_0$ from Eq. (\ref{equm3}),
\begin{equation}
\boldsymbol{y}_0=\frac{1}{\sqrt{\gamma_t}} \left(\boldsymbol{y}_t+\sqrt{1-\gamma_t}f_\theta\left(\boldsymbol{y}_c, \boldsymbol{y}_t^*, \gamma_t\right)\right)
\label{equm5}
\end{equation}\par 

After generating the approximate $\boldsymbol{y}_0$, we can estimate the parameterized mean of $p_\theta(\boldsymbol{y}_{t-1}\mid\boldsymbol{y}_t, \boldsymbol{y}_c)$ by substituting the estimated $\boldsymbol{y}_0$ into the posterior distribution $q(\boldsymbol{y}_{t-1} \mid \boldsymbol{y}_0, \boldsymbol{y}_t)$ in Eq. (\ref{equm2}). With the parameterization trick, the state for the last time $t-1$ in the reverse process can be sampled,
\begin{equation}
\boldsymbol{y}_{t-1}=\sqrt{\gamma_{t-1}}\left(\boldsymbol{y}_t-\frac{\sqrt{1-\gamma_t}}{\sqrt{\gamma_t}} f_\theta\right)+\sqrt{1-\gamma_{t-1}-\sigma_t^2} f_\theta+\sigma_t \boldsymbol{z}, \boldsymbol{z} \sim \mathcal{N}(0, \mathbf{I})
\label{equm6}
\end{equation}
Here, we set $\sigma_t$ to 0 by setting $\eta = 0$, such that the randomness induced by the reverse sampling is eliminated. Moreover, this configuration enables the respacing trick to accelerate the sampling process, training a model with an arbitrary number of forward steps but only sampling from some of them \cite{song_denoising_2022}. In our experiment, we use quadratic sampling with 500 steps for the 2000 steps diffusion model, while the 100-step model does not use the respacing trick. $\boldsymbol{y}_c$ is set to the raw pattern corresponding to $\boldsymbol{y}_0$.

\subsection*{ControlNet Implementation}\label{secm6}
ControlNet injects additional conditions into the blocks of a Palette network, where the condition is provided by the output of IRAE. The implementation of ControlNet requires pre-training the network of Palette, and then fixing the pre-training weights to perform the additional control step for the noise prediction. Before the implementation of the control step, the additional single channel condition $\hat{\boldsymbol{y}}_c$ is transformed by a trainable encoder, containing 7 layers 3 $\times$ 3 convolution and sigmoid activation unit (channels are 16, 16, 32, 32, 96, 96 and 256, respectively), and use a 3 $\times$ 3 convolution to compress 256 channels to one as a constraint pattern in ControlNet. The next control step is cloning the network block of Palette to a trainable copy with parameters, and taking the external constraint pattern as input, which is the same as in the traditional ControlNet in Ref. \cite{zhang_adding_2023}. The IRAE's output is selected as $\hat{\boldsymbol{y}}_c$.

\subsection*{General diffusion model and DDM}\label{secm7}
Standard DPMs are built based on a forward process to degrade an original image to a contaminated pattern with Gaussian noise, and a trained restoration network to perform the denoising until a clear image is generated. However, the general diffusion considers arbitrary degradation operations rather than mere Gaussian noise, which can use a randomized or deterministic degeneration function \cite{bansal_cold_2023}. Under the general framework, the degradation operation is defined as $D(\boldsymbol{y}_0, t)$ and the restoration operation $R(\boldsymbol{y}_t, t)$, in which we set $D(\boldsymbol{y}_0, t)=\boldsymbol{y}_t$ and $R(\boldsymbol{y}_t, t)=\boldsymbol{y}_0$. Resembling the standard DPMs, the general diffusion model employs the neural network to learn the parameterized restoration operation $R_\theta(\boldsymbol{y}_t, t)$ by minimizing the formulation,
\begin{equation}
\underset{\theta}{\arg \min } \mathbb{E}_{\boldsymbol{x}}\left\|\mathcal{L}\left(R_\theta(D(\boldsymbol{x}, t), t), \boldsymbol{x}\right)\right\|
\label{equm7}
\end{equation}
where $\mathcal{L}$ is a common loss function, e.g., mean absolute error, and $\boldsymbol{y}_0 = \boldsymbol{x}$.\par 

In order to achieve the deterministic diffusion, the degeneration function D(y0,t) needs to satisfy $D(\boldsymbol{y}_0, 0)=\boldsymbol{x}$ and $D(\boldsymbol{y}_0, T)=\boldsymbol{y}_T$, where $\boldsymbol{x}$ and $\boldsymbol{y}_T$ are the ground truth and the generated raw pattern in the experiment, respectively. Once these conditions are met, the randomness of the final state in the forward process is removed. When the deterministic degenerate function is further selected, the randomness of the diffusion path can also be eliminated. Although the degradation function still has various choices, we employ a rectified straight flow in our DDM,
\begin{equation}
    D(\boldsymbol{x},t)=\alpha_t\boldsymbol{x}+(1-\alpha_t)\boldsymbol{y}_T
    \label{equm8}
\end{equation}
where $\alpha_t$ is a weight associated with time $t$. The $\alpha_t$ can be both predefined and learned by reparameterization. In our practice, we predefine the $\alpha_t = \mathrm{cos}^2(\pi t/2T)$. There are three advantages to using Eq. (\ref{equm8}). Firstly, the linear function enables the straight diffusion path, allowing the training and sampling process with a few diffusion steps to accelerate the training and sampling process \cite{liu_flow_2022}. Secondly, the degradation function maintains the pixel range of the degraded pattern at an arbitrary time $t$ that is consistent with $\boldsymbol{x}$ and $\boldsymbol{y}$, stabilizing the training. Thirdly, the simple form of Eq. (\ref{equm8}) is effortless to differentiate, which is important for the modeling of uncertainty quantification. \par 

For the sampling process, when the neural network is trained to implement the restoration function, the sampling can directly use $D(R_\theta(\boldsymbol{y}_t, t), t-1)$ to infer the $\boldsymbol{y}_{t-1}$, the same as the standard DPMs. However, the neural network is invariably imperfect, making the direct inference inaccurate, especially in the DDM. To overcome this problem, we employ the indirect sampling algorithm to achieve high quality results under the imperfect restoration. This superior sampling algorithm is \cite{bansal_cold_2023},
\begin{equation}
    \boldsymbol{y}_{t-1}=\boldsymbol{y}_t-D(R_\theta(\boldsymbol{y}_t, t), t)+D(R_\theta(\boldsymbol{y}_t, t), t-1)
    \label{equm9}
\end{equation}
which is extremely tolerant for the restoration error, providing a stable reverse path. Using this algorithm, image reconstruction can be completed in arbitrary CI tasks by DDM. We experimentally compare the two sampling algorithms and verify the significant advantage of indirect sampling, shown in Extended Data Fig. \ref{figed4}.

\subsection*{The uncertainty quantification in DDM}\label{secm8}
To capture the uncertainty, the BNN uses probability distributions to replace the deterministic network weights and to approximate Bayesian inference in deep Gaussian processes \cite{feng_deep-learning-based_2021}. Assuming the training dataset$\left( \boldsymbol{Y},\boldsymbol{X} \right)=\left\{ \boldsymbol{y}_{t}^{n},{\boldsymbol{x}^{n}} \right\}_{n=1}^{N}$ with a size of $N$, the predictive distribution $p\left(\boldsymbol{x}^* \mid \boldsymbol{y}_t^*, \boldsymbol{X}, \boldsymbol{Y}\right)$ can be modeled by the Bayes' theorem,
\begin{equation}
p\left(\boldsymbol{x}^* \mid \boldsymbol{y}_t^*, \boldsymbol{X}, \boldsymbol{Y}\right)=\int p\left(\boldsymbol{x}^* \mid \boldsymbol{y}_t^*, \boldsymbol{w}\right) p(\boldsymbol{w} \mid \boldsymbol{X}, \boldsymbol{Y}) d \boldsymbol{w}
\label{equm10}
\end{equation}
where all possible network weights $\boldsymbol{w}$ is marginalized and $\boldsymbol{y}_t^*$ is a given test input. Eq. (\ref{equm10}) indicates that the predictive distribution is related to the probability, $p\left(\boldsymbol{x}^* \mid \boldsymbol{y}_t^*, \boldsymbol{w}\right)$, of the output $\boldsymbol{x}^*$ given the input $\boldsymbol{y}_t^*$ and the weights $\boldsymbol{w}$, and the probability, $p(\boldsymbol{w} \mid \boldsymbol{X}, \boldsymbol{Y})$, of the weights $\boldsymbol{w}$ given the training data $(\boldsymbol{X}, \boldsymbol{Y})$. By modeling the two probabilities, the model and data uncertainty can be quantified, respectively.\par 

For image data with total $K$ pixels that obey a Gaussian distribution for each pixel, the probability $p\left(\boldsymbol{x}^* \mid \boldsymbol{y}_t^*, \boldsymbol{w}\right)$ is,
\begin{equation}
p\left(\boldsymbol{x}^* \mid \boldsymbol{y}_t^*, \boldsymbol{w}\right)=\prod_{k=1}^K \frac{1}{\sqrt{2 \pi} \sigma^k} \exp \left[-\frac{\left(x^{*k}-\mu^k\right)^2}{2\left(\sigma^k\right)^2}\right]
\label{equm11}
\end{equation}
where the $\mu^k$ and $\sigma^k$ are the mean and standard deviations for the $k$-th pixel, respectively. It means that the data uncertainty can be captured when the training process minimizes the negative log-likelihood function $\mathcal{L} = -\mathrm{log}\left(p\left(\boldsymbol{x}^* \mid \boldsymbol{y}_t^*, \boldsymbol{w}\right)\right)/K$. Note that $\boldsymbol{x}^*$ is the ground truth, $\boldsymbol{\mu}$ and $\boldsymbol{\sigma}$ are the predicted results of BNN. For model uncertainty quantification, it requires to estimate the probability $p(\boldsymbol{w} \mid \boldsymbol{X}, \boldsymbol{Y}) d \boldsymbol{w}$. Generally, a simple distribution $q(\boldsymbol{w})$ can be learned to approximate the posterior $p(\boldsymbol{w}\mid\boldsymbol{X}, \boldsymbol{Y}) d \boldsymbol{w}$ by the dropout network which applies learnable concrete dropout before every weight layer. \par 

By Monte Carlo sampling $\boldsymbol{w}$ from $q(\boldsymbol{w})$, we can approximately calculate Eq. (\ref{equm10}),
\begin{equation}
p\left(\boldsymbol{x}^* \mid \boldsymbol{y}_t^*, \boldsymbol{X}, \boldsymbol{Y}\right) \approx \int p\left(\boldsymbol{x}^* \mid \boldsymbol{y}_t^*, \boldsymbol{w}\right) q(\boldsymbol{w}) d \boldsymbol{w} \approx \frac{1}{S} \sum_{s=1}^S p\left(\boldsymbol{x}^* \mid \boldsymbol{y}_t^*, \boldsymbol{w}^{(s)}\right)
\label{equm12}
\end{equation}
When dropout is activated and kept at the inference stage, the input data undergoes stochastic forward propagation with the trained BNN, during which we estimate $p\left(\boldsymbol{x}^* \mid \boldsymbol{y}_t^*, \boldsymbol{X}, \boldsymbol{Y}\right)$ by averaging $S$ samplings by using Eq. (\ref{equm12}). The predictive mean can also be computed as the prediction of the BNN according to Eq. (\ref{equm12}),
\begin{equation}
\hat{\boldsymbol{\mu}}_t=\mathbb{E}\left(\boldsymbol{x}^* \mid \boldsymbol{y}_t^*, \boldsymbol{X}, \boldsymbol{Y}\right) \approx \frac{1}{S} \sum_{s=1}^S \boldsymbol{\mu}^{(s)}
\label{equm13}
\end{equation}
Under this condition, the model and data uncertainty can be quantified,

\begin{equation}
\begin{aligned}
    \boldsymbol{\sigma}_t^{(M)}=\sqrt{\mathbb{E}\left(\operatorname{Var}\left(\boldsymbol{x}^* \mid \boldsymbol{y}_t^*, \boldsymbol{X}, \boldsymbol{Y}\right)\right)} \approx \sqrt{\frac{1}{S} \sum_{s=1}^s\left(\boldsymbol{\mu}^{(s)}-\hat{\boldsymbol{\mu}}_t\right)^2} \\
    \boldsymbol{\sigma}_t^{(D)}=\sqrt{\operatorname{Var}\left(\mathbb{E}\left(\boldsymbol{x}^* \mid \boldsymbol{y}_t^*, \boldsymbol{X}, \boldsymbol{Y}\right)\right)} \approx \sqrt{\frac{1}{S} \sum_{s=1}^S\left(\boldsymbol{\sigma}^{(s)}\right)^2}
\end{aligned}
\label{equm14}
\end{equation}\par 

Eq. (\ref{equm13}) and Eq. (\ref{equm14}) quantify the mean and uncertainty in a single step network. In our DDM with multistep operations, the final data/model uncertainty requires the propagation of uncertainty. Considering the Eq. (\ref{equm8}) and Eq. (\ref{equm9}), the propagation of uncertainty is,
\begin{equation}
\begin{aligned}
\left(\boldsymbol{\sigma}^{(M / D)}\right)^2=&\sum_{t=1}^T\left[\left(\alpha_t-\alpha_{t-1}\right) \boldsymbol{\sigma}_t^{(M / D)}\right]^2\\
&+2 \sum_{t=1}^T\left[\left(\alpha_t-\alpha_{t-1}\right) \operatorname{cov}\left(\boldsymbol{y}_t, R_\theta\left(\boldsymbol{y}_t, t\right)\right)^{(M / D)}\right]
\end{aligned}
\label{equm15}
\end{equation}
When variables are independent, the covariance is zero in Eq. (\ref{equm15}). The first term represents the respective contributions of each variable to the total uncertainty, while the second term measures the overall error of two variables. In DDM, the degraded pattern at the current time $t$ is associated with that at the $t-1$ time, such that the uncertainty is also associated by the covariance. However, calculating the covariance is extremely complex, especially when the relationship between variables is nonanalytic. In this case, we introduce statistical techniques and Monte Carlo simulations to estimate the covariance. The two covariances have different forms,
\addtocounter{equation}{1}
\begin{equation}
\operatorname{cov}\left(\boldsymbol{y}_t, R_\theta\left(\boldsymbol{y}_t, t\right)\right)^{(M)} \approx \frac{1}{H} \sum_{h=1}^H\left(\boldsymbol{\mu}_{t+1}^{(h)}-\hat{\boldsymbol{\mu}}_{t+1}\right) \cdot\left(\boldsymbol{\mu}_t\left(\boldsymbol{\mu}_{t+1}^{(h)}\right)-\hat{\boldsymbol{\mu}}_t\left(\boldsymbol{\mu}_{t+1}\right)\right) \tag{\theequation a}\label{equm16a}
\end{equation}
\begin{equation}
\operatorname{cov}\left(\boldsymbol{y}_t, R_\theta\left(\boldsymbol{y}_t, t\right)\right)^{(D)} \approx \frac{1}{H} \sum_{h=1}^H\left(\boldsymbol{\sigma}_{t+1}^{(h)}\right) \cdot\left(\boldsymbol{\sigma}_t\left(\boldsymbol{\mu}_{t+1}^{(h)}\right)\right)\tag{\theequation b}\label{equm16b}
\end{equation}
where the covariance is zero at the $T$-th diffusion step. $\hat{\boldsymbol{\mu}}_t\left(\boldsymbol{\mu}_{t+1}^{(h)}\right)$ represents the average of the outputs when $\boldsymbol{\mu}_{t+1}$ is used as input and is calculated by Eq. (\ref{equm13}). Here, we assume that the correlation coefficient is 1 in Eq. (\ref{equm16b}) because the data uncertainty is consistently increasing. The covariance of data uncertainty also can be calculated by sampling average $\mathbb{E}_{\boldsymbol{w}}\left[\mathbb{E}_{\boldsymbol{y}_{t+1}}\left[\boldsymbol{y}_{t+1} \cdot \boldsymbol{\mu}_t\left(\boldsymbol{y}_{t+1}\right)\right]-\boldsymbol{\mu}_{t+1} \mathbb{E}_{\boldsymbol{\mu}_t}\left[\boldsymbol{\mu}_t\left(\boldsymbol{y}_{t+1}\right)\right]\right]$, where the $\boldsymbol{y}_{t+1}$ is generated by sampling from the Gaussian distribution $\mathcal{N}(\boldsymbol{\mu}_{t+1},\boldsymbol{\sigma}_{t+1})$. Experimentally, $H$ is set to 24.\par

\subsection*{Network Implementation and computational resource}\label{secm9}
DDM employs the network architecture in Ref. \cite{bansal_cold_2023} as the backbone network, which resembles the traditional DPMs \cite{ho_denoising_2020}. The concrete dropout module is embedded into the backbone network to build the BNN. Different from placing dropout before every weight layer, we introduce a concrete dropout before each encoder/decoder block, thereby reducing the consumption of computing power and memory but maintaining the accuracy of the uncertainty quantification. All network models are built and implemented with the PyTorch framework \cite{paszke_pytorch_2019}. The dataset is split with 90\% for training and 10\% for testing. The Adam optimizer with dynamic learning rates is applied to optimize the network parameters. The focal frequency loss \cite{jiang_focal_2021} is used as the loss function for CNN and IRAE, and for DDM without uncertainty prediction. Palette and ControlNet employ the mean absolute error, while the DDM with uncertainty prediction uses the negative log-likelihood function, as the loss function. CNN and IRAE are trained and tested on a single NVIDIA V100 32GB professional graphics processing unit (GPU), Palette and ControlNet are 4$\times$V100 32GB, while the DDM uses a single NVIDIA RTX4090 24GB consumer-grade GPU (keeping data accuracy on TensorFloat32).

\backmatter

\bmhead{Data availability}
All data used to produce the findings of this study are available from the corresponding author on reasonable request.

\bmhead{Code availability}
Codes for the whole pipeline of this study are available via GitHub at \url{https://github.com/weirufan/DDM}.

\bmhead{Acknowledgements}
This work was supported by the National Natural Science Foundation of China (Grant No. 11934011, 62075194, U21A6006, 62202418, U21B2004), National Key Research and Development Program of China (Grant No. 2019YFA0308100, 2023YFB2806000, 2022YFA1204700), the Strategic Priority Research Program of Chinese Academy of Sciences (Grant No. XDB28000000), the Open Program of the State Key Laboratory of Advanced Optical Communication Systems and Networks at Shanghai Jiao Tong University (Grant No. 2023GZKF024), the Fundamental Research Funds for the Central Universities, the Information Technology Center and State Key Lab of CAD\&CG, and the Zhejiang Provincial Key Laboratory of Information Processing, Communication and Networking (IPCAN).

\bmhead{Author contributions}
W.F. and Y.L. conceived the idea and designed the experiment. W.F. carried out the experiment, collected data and performed the theoretical modeling of DDM. W.F. wrote the control programs for devices and the code of DDM. X.T. wrote and implemented the Palette and ControlNet. W.F., Y.L. and D.W.W. analyzed data and wrote the manuscript. The project is supervised under D.W.W. All authors discussed the results and revised the manuscript.


\newpage
\begin{appendices}
\captionsetup[figure]{labelfont={bf},labelformat={default},labelsep=period,name={Extended Data Fig.}}

\begin{figure}[hbt]
\centering
\includegraphics[width=\textwidth]{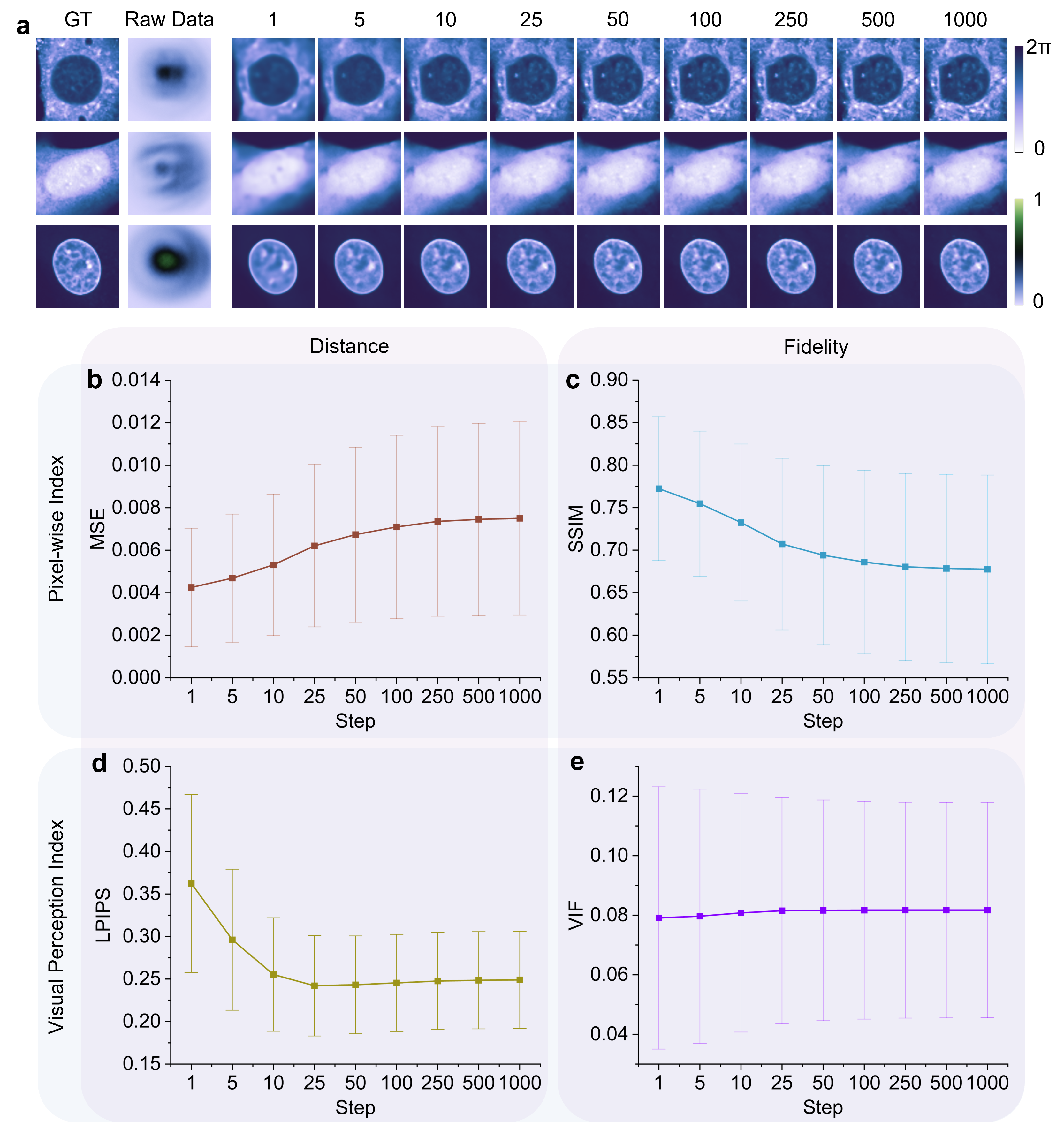}
\caption{The sampling algorithm of DDM. The top line shows reconstructed results by the direct sampling algorithm. In the reverse process, the reconstruction deviation progressively accumulates until the image is completely drowned in the noise. The results of the indirect sampling algorithm are shown on the bottom line. It is robust in progressive recovering the image, demonstrating the superiority of deterministic diffusion.}\label{figed1}
\end{figure}

\newpage

\begin{figure}[hbt]
\centering
\includegraphics[width=\textwidth]{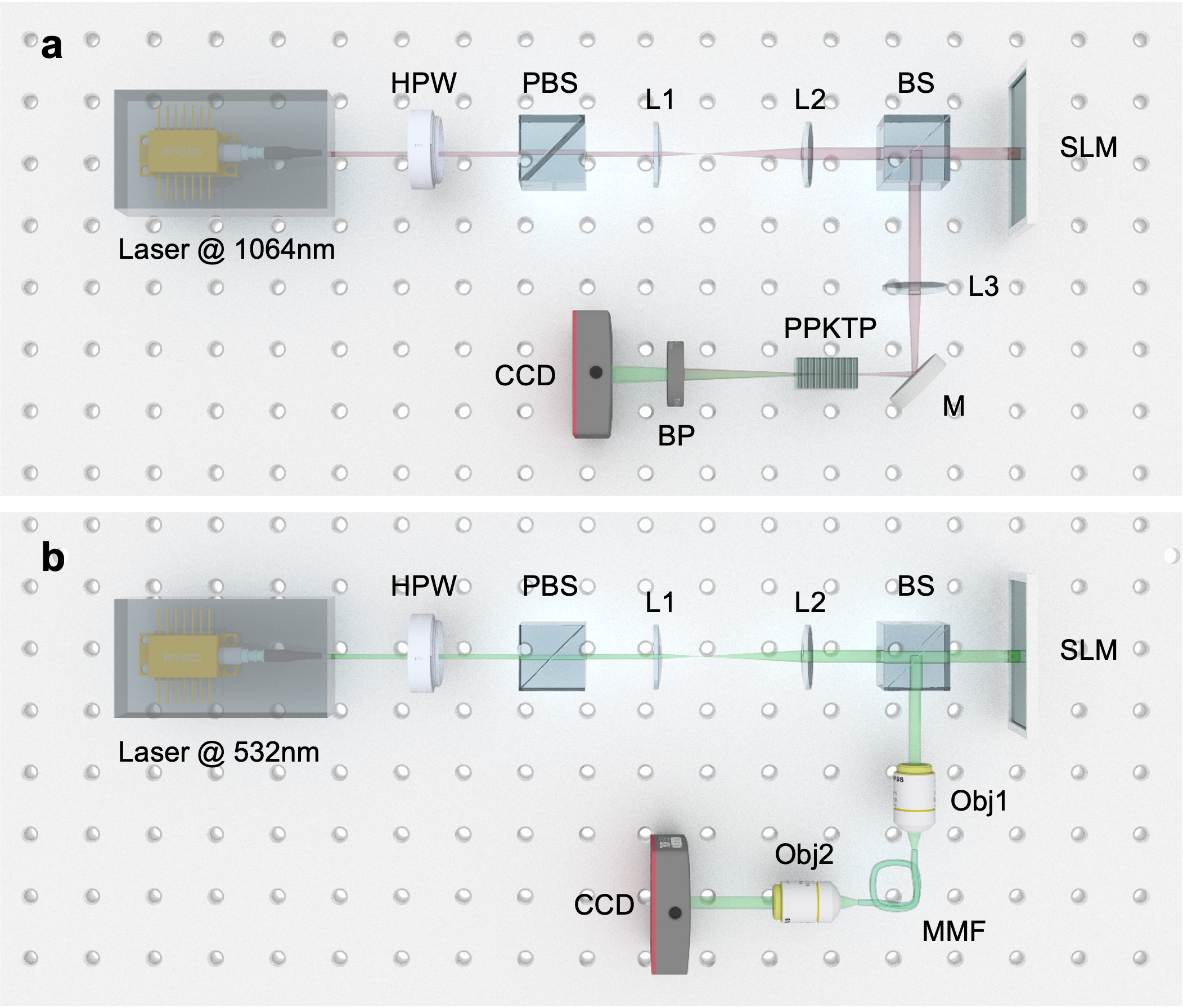}
\caption{Experimental setup. \textbf{a,} and \textbf{b,} Generating the nonlocal pattern by SHG process and the scattering of MMF. HWP: half-wave plate; PBS: polarized beam splitter; L: lens; BS: non-polarized beam splitter; M: mirror; BP: bandpass filter; Obj: objective; SLM: spatial light modulator; CCD: charge coupling device.}\label{figed2}
\end{figure}

\newpage

\begin{figure}[hbt]
\centering
\includegraphics[width=\textwidth]{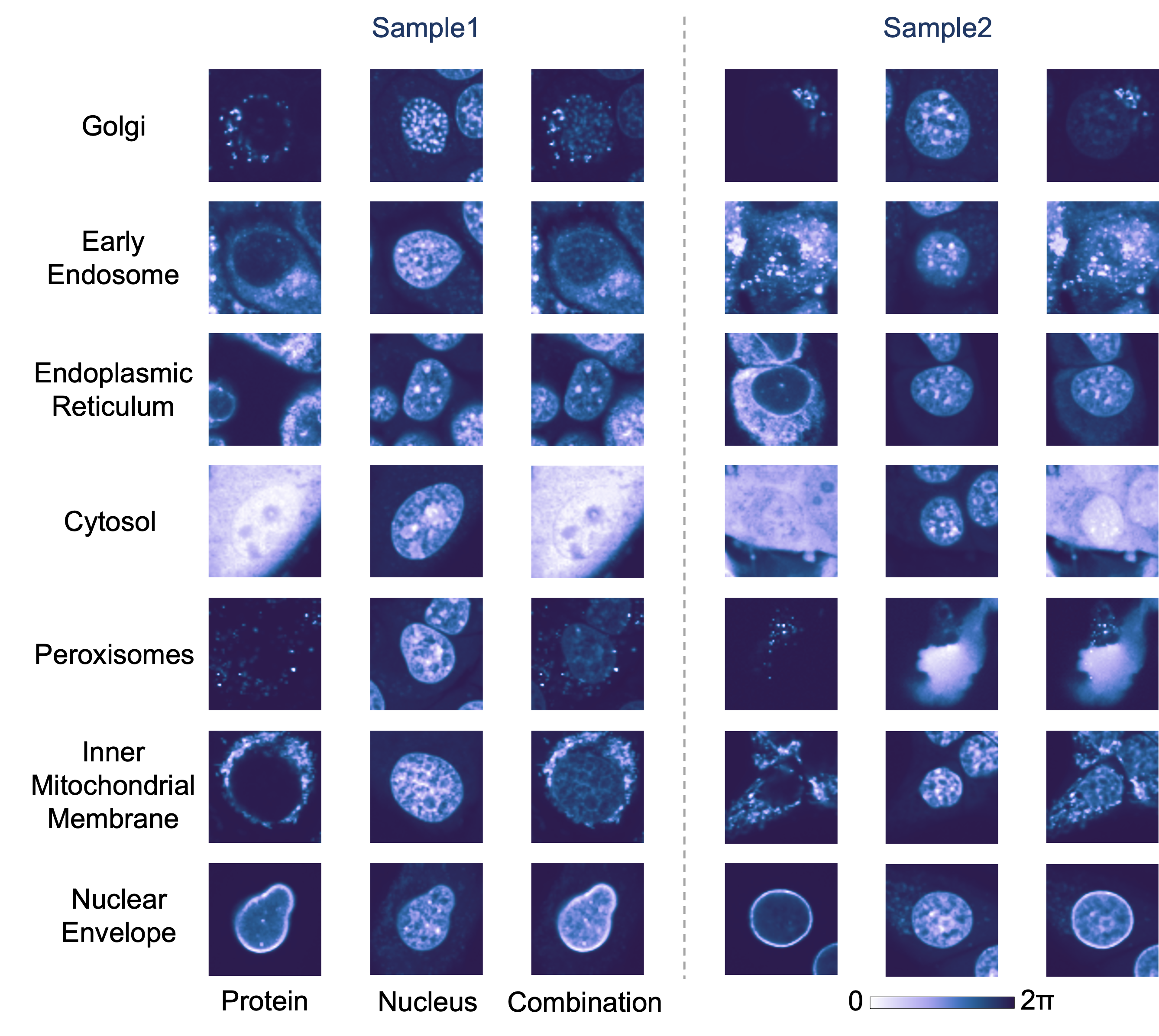}
\caption{Samples of dataset. The original data come from the COOS-7 dataset. The nucleus has a similar structure in appearance, but the protein is significantly different. We combine the protein and nucleus to generate a structurally complex image.}\label{figed3}
\end{figure}

\newpage

\begin{figure}[hbt]
\centering
\includegraphics[width=\textwidth]{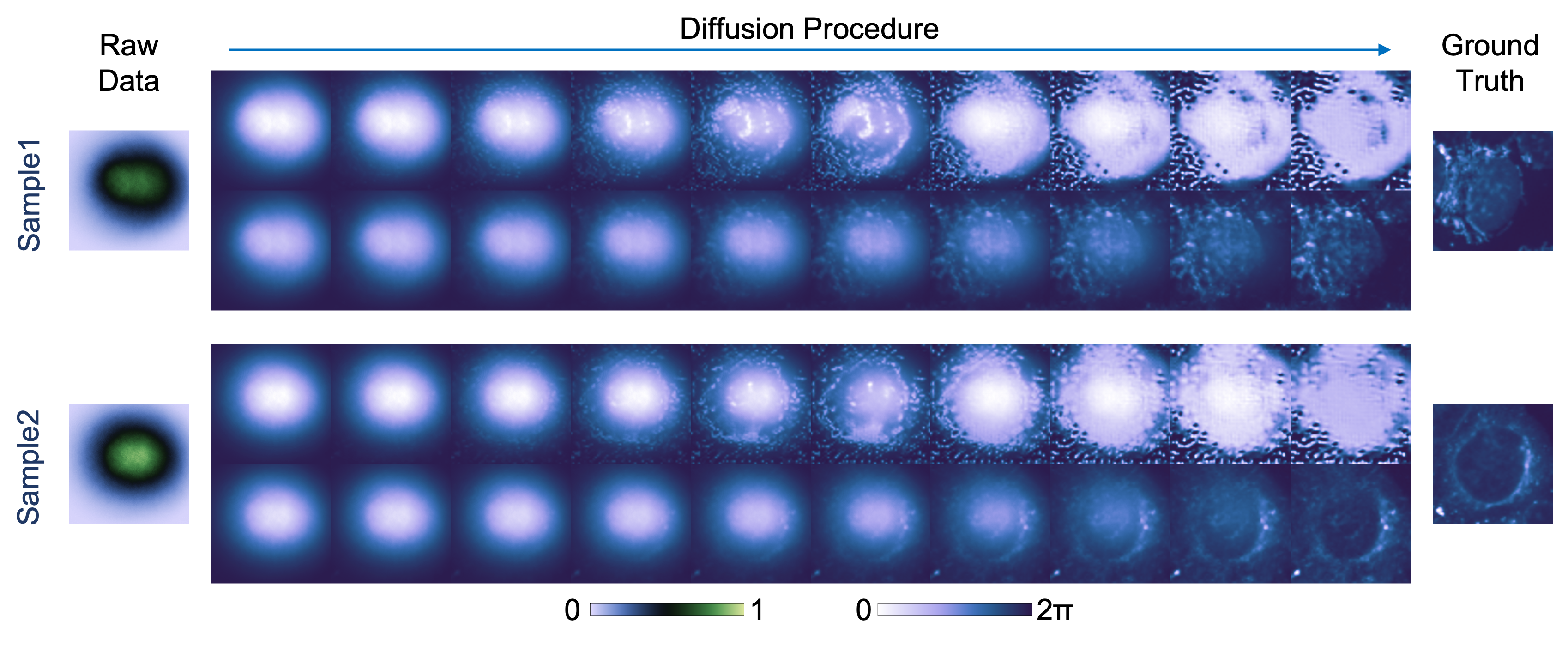}
\caption{The results of difference diffusion steps. \textbf{a,} Some reconstructed results. The more diffusion steps there are, the more conservative and smoother the results product. \textbf{b}-\textbf{e} are the metrics of image change with the diffusion step. MSE: mean squared error (lower value indicates smaller error); SSIM: structure similarity index measure (larger value indicates higher similarity); LPIPS: learned perceptual image patch similarity (lower value indicates higher similarity); VIF: vision information fidelity (larger value indicates higher fidelity). The pixel-level errors (MSE) and fidelities (SSIM) in smaller diffusion step have a smaller average value and variance, but in visional perception counterparts (LPIPS and VIF) have a lower score and lager variance. Performance changes tend to saturate as the steps increase to about 250. Error bar: $\pm$ standard deviation.}\label{figed4}
\end{figure}
\end{appendices}

\end{document}